\documentstyle[eqsecnum,aps]{revtex}

\begin{document}
\draft
\title{ Screw instability in black hole magnetospheres \\
        and a stabilizing effect of field-line rotation}
\author{Akira Tomimatsu and Tomoya  Matsuoka}
\address{Department of Physics, Nagoya University, Chikusa-ku, 
         Nagoya 464-8602, Japan}
\author{Masaaki Takahashi}
\address{Department of Physics and Astronomy, Aichi University of 
         Education, \\ Kariya, Aichi 448-8542, Japan}

\maketitle

\begin{abstract}
The screw instability of magnetic field is a mechanism for prohibiting a 
generation of strongly twisted field lines in large scales. If it can work 
in black hole magnetospheres, the global axisymmetric structure and the 
main process of energy release will be significantly influenced. In this 
paper, we study the instability condition in the framework of the 
variational principle, paying special attention to a stabilizing effect due 
to field-line rotation against the screw-unstable modes satisfying the 
well-known Kruskal-Shafranov criterion. The basic formulation for the 
stability analysis of rotating force-free fields in Kerr geometry is 
provided. Then, for the practical use of the analytical method, the 
stationary configuration is assumed to be cylindrical, and we treat the 
long-wave mode perturbations excited at large distances from the central 
black hole, where the strength of gravity becomes negligibly small. This 
allows us to derive clearly the new criterion dependent on the rotational 
angular velocity of magnetic field lines, and the implications of the 
results obtained for magnetic activities of jets driven by a black hole are 
briefly discussed.
\end{abstract}



\section{INTRODUCTION}
\label{sec:level1}

The global structure of black hole magnetospheres involving axisymmetric 
magnetic field and plasma injected from an accretion disk has been 
extensively investigated for explaining various observational features of 
active galactic nuclei (AGNs) (see, e.g., \cite{BES97} for a review). The 
magnetic field lines which are assumed to be frozen to the rotating plasma 
may be twisted and rotating, which means that the components of toroidal 
magnetic field (i.e., poloidal electric currents) and poloidal electric 
field are generated. The magnetic tension of the toroidal field can play an 
important role of forming cylindrically collimated jets over significantly 
long distances, and if the magnetic field lines thread a rotating black 
hole in the central region of AGN, the so-called Blandford-Znajek mechanism 
\cite{TPM86,BZ77} is believed to become a possible process for powering the 
plasma jets through Poynting flux.

However, it is also well-known from the field of plasma physics that the 
magnetic field configurations with both poloidal component and toroidal one 
can be screw-unstable \cite{kad66,bat78}, and this process of a current 
driven instability makes the magnetic field lines untwist on short 
time-scales and causes a sudden release of magnetic free-energy to plasma's 
kinetic energy. In conventional plasma confinements, the screw instability 
is claimed to work for long-wave mode perturbations, if the equilibrium 
magnetic field has a toroidal component exceeding the limit given by the 
Kruskal-Shafranov criterion. Then, according to this instability condition 
one may discuss that the power of the Blandford-Znajek mechanism which is 
crucially dependent on the strength of of toroidal magnetic field is 
significantly suppressed \cite{li00}, and the mechanism for acceleration 
and collimation of jets also becomes less efficient. Further, a flare-like 
energy radiation is expected to occur in the screw-unstable magnetospheres 
\cite{gru99}. After a part of the magnetic energy is released through the 
process of the screw instability, the magnetic field lines threading a 
rotating black hole may be twisted again for generating the toroidal 
component stronger than the Kruskal-Shafranov threshold, and the flaring 
may be quasi-periodically observed.

Thus, the screw instability is an astrophysically interesting process which 
can significantly influence the global axisymmetric structure and the main 
process of energy release in black hole magnetospheres. For the application 
to the astrophysical problems, however, the instability condition should be 
investigated in more detail. In particular, an essential feature of black 
hole magnetospheres is that the twisted magnetic field lines are rotating. 
In this paper, our main purpose is to consider the effect of the field-line 
rotation in the stability analysis of stationary magnetic configurations 
and to derive the new criterion of the screw instability. In Sec. II, in 
the framework of the force-free approximation in Kerr geometry, we give the 
basic equations for the axisymmetric stationary fields and the 
non-axisymmetric linear perturbations, following the usual procedure to 
derive the variational principle \cite{kad66,kam67}. For the practical use 
of the analytical method, in Sec. III, the stationary configuration is 
assumed to be cylindrical, and we treat the long-wave mode perturbations 
excited at large distances from the central black hole, where the strength 
of gravity becomes negligibly small. Then we can clearly see that the 
field-line rotation has a stabilizing effect against the screw-unstable 
modes satisfying the Kruskal-Shafranov criterion. This allows us to 
reconsider the conclusions based on the Kruskal-Shafranov criterion as a 
first step toward more complete investigations using the fully general 
relativistic formulae given in Sec. II. While any direct influences of Kerr 
geometry are missed in the new criterion obtained here (which is relevant 
to force-free rotating magnetospheres far distant from various central 
objects), it is possible to discuss the implications for magnetic 
activities in black hole magnetospheres, by estimating the toroidal 
component of the stationary magnetic field under the assumption that it is 
generated as a result of the black hole rotation. Section IV is devoted to 
a brief comment on this problem. Throughout this paper, we use the 
geometric units with $G=c=1$.

\section{STABILITY ANALYSIS IN KERR GEOMETRY}
\label{sec:level2}

In the framework of the 3+1 formalism in Kerr geometry we study a current 
driven instability of electromagnetic fields under the force-free 
approximation which will be valid in black hole magnetospheres except near 
the event horizon and the (inner and outer) light cylinders \cite{pun91}. 
The Kerr metric is written in the Boyer-Lindquist coordinates as follows,
\begin{equation}
   ds^2 = - \alpha ^2 dt^2 + h_{i j} (d x^j - \beta^j dt)(dx^i - \beta^i dt) ,
\end{equation}
where
\begin{eqnarray}
   \alpha = \frac{\rho}{\Sigma} \sqrt{\Delta} , \ \
   \beta^r = \beta^{\theta} = 0  , \ \
   \beta^{\phi} =  \frac{2 M a r}{\Sigma^2} \equiv  \omega , \\
   h_{rr} = \frac{\rho^2}{\Delta} , \ \
   h_{\theta \theta} = \rho^2  ,  \ \
   h_{\phi \phi} = \varpi^2, \\
   \rho^2 = r^2 + a^2 \cos^2 \theta  , \ \
   \Delta = r^2 + a^2 - 2 M r , \\
   \Sigma^2 = (r^2 + a^2)^2 - a^2 \Delta \sin^2 \theta  , \ \
   \varpi = \frac{\Sigma}{\rho} \sin \theta,
\end{eqnarray}
and $M$ and $a$ denote the mass and angular momentum per unit mass of the 
black hole, respectively. Then, the Maxwell equations and the force-free 
condition for zero-angular-momentum observers (ZAMOs) are given by
\begin{eqnarray}
(\partial_t + L_{\beta}) \vec{B} &=& - \nabla \times  \alpha \vec{E} , 
\label{maxwell1} \\
(\partial_t + L_{\beta}) \vec{E} &=& \nabla \times  \alpha \vec{B} - \alpha 
\vec{j} , \label{maxwell2} \\
\rho \vec{E} + \vec{j} \times \vec{B} &=& 0 , \label{ff}
\end{eqnarray}
where $L_{\beta}$ denotes the Lie derivative along $\beta^{i}$, $\vec{B}$ 
and $\vec{E}$ are the magnetic and electric fields, $\rho$ and $\vec{j}$ 
are the electric charge and the current densities multiplied by $4 \pi$ 
\cite{TPM86}.

If the current and charge densities are eliminated from the Maxwell 
equations, we have the equation
\begin{equation}
\partial_t (\vec{B} \times \vec{E}) + \vec{B} \times L_{\beta} \vec{E}+ 
(L_{\beta} \vec{B}) \times \vec{E} + \alpha (\vec{\nabla} \cdot \vec{E}) 
\vec{E}
= \vec{B} \times (\vec{\nabla} \times  \alpha \vec{B}) + \vec{E} \times 
(\vec{\nabla} \times  \alpha \vec{E}) . \label{ii}
\end{equation}
Because the force-free fields may be regarded as a magnetically dominated 
limit in magnetohydrodynamics (MHD), we introduce the plasma velocity 
defined by
\begin{equation}
  \vec{E} + \vec{v} \times \vec{B} = 0 , \label{v}
\end{equation}
and we have the frozen-in law written by
\begin{equation}
(\partial_t + L_{\beta})\vec{B} = \vec{\nabla} \times  (\vec{v} \times 
\alpha \vec{B}) . \label{i}
\end{equation}
These are the basic set of dynamical equations for the force-free fields.

For axisymmetric stationary configurations of electromagnetic and velocity 
fields denoted by $\vec{B}_{0}$, $\vec{E}_{0}$ and $\vec{v}_{0}$, we obtain 
the well-known relations. For example, the poloidal component 
$\vec{B}^{p}_{0}$ of the magnetic field is written by the magnetic stream 
function $\Psi$ as follows,
\begin{equation}
\vec{B}_{0}^{p} = \frac{\nabla\Psi \times \vec{e}_{T}}{2\pi\varpi} ,
\end{equation}
where $\vec{e}_{T}=(1/\varpi)(\partial/\partial\phi)$ means the toroidal 
basis vector which has unit norm. In stationary rotating magnetospheres the 
velocity field is given by
\begin{equation}
\vec{v}_0 \equiv v_0^{T} \vec{e}_{T} = (\varpi/\alpha) (\Omega_F - \omega) 
\vec{e}_{T} ,
\end{equation}
where $\Omega_{F}=\Omega_{F}(\Psi)$ is constant along a magnetic field line 
of $\Psi(r,\theta)=$ constant, and we call it the angular velocity of a 
magnetic field line. This rotational motion generates the poloidal electric 
field $\vec{E}_{0} = -\vec{v}_{0} \times \vec{B}_{0}$. Further, if the 
magnetic field lines are twisted, the poloidal current $I=I(\Psi)$ (which 
is chosen to be positive for downward flows) is generated according to the 
equation
\begin{equation}
I = - \alpha \varpi B_{0}^{T}/2 ,
\end{equation}
where $B_{0}^{T}$ is the toroidal magnetic component. Then, Eq. (\ref{ii}) 
reduces to the Grad-Shafranov equation for $\Psi$ \cite{fro98},
\begin{equation}
 \frac{\Omega_F - \omega}{\alpha}(\vec{\nabla} \Psi)^{2}
 \frac{d \Omega_F}{d\Psi}
+ \vec{\nabla} \cdot \left\{ \frac{\alpha}{\varpi^2}
 \left[1 - \left(\frac{\Omega_F - \omega}{\alpha}\varpi \right)^2 \right]
 \vec{\nabla} \Psi \right\}
+ \frac{16 \pi^2}{\alpha \varpi^2}I\frac{dI}{d\Psi} = 0 , \label{GS}
\end{equation}
which should be solved by giving the functions $I(\Psi)$ and 
$\Omega_F(\Psi)$ from boundary conditions at a plasma source. Though we do 
not try to solve the highly nonlinear equation (\ref{GS}) for background 
fields, it will be used for simplifying the equations for linear 
perturbations in the stability analysis in the following.

Now, we consider small perturbations $\vec{B}_{1}$, $\vec{E}_{1}$ and 
$\vec{v}_{1}$ from the stationary configurations. It is easy to derive the 
linearized versions of Eqs. (\ref{ii}), (\ref{v}) and (\ref{i}) as follows,
\begin{eqnarray}
\vec{E}_1 &=& - \vec{v}_0 \times \vec{B}_1 - \vec{v}_1 \times \vec{B}_0 , 
\label{v'} \\
(\partial_t + L_{\beta})\vec{B}_1 &=& \vec{\nabla} \times 
(\vec{v}_1 \times \alpha \vec{B}_0 + \vec{v}_0 \times \alpha \vec{B}_{1}) , 
\label{i'} \\
\vec{B}_0 \times \partial_t \vec{E}_1 + \vec{B}_0 \times L_{\beta} 
\vec{E}_1 &+& \vec{B}_1 \times L_{\beta} \vec{E}_0
+ \alpha \left[(\nabla \cdot \vec{E}_0)\vec{E}_1 
+ (\nabla \cdot \vec{E}_{1})\vec{E}_{0} \right]         \nonumber \\
&=&  \vec{B}_0 \times (\vec{\nabla} \times \alpha \vec{B}_{1}) + \vec{B}_1 
\times (\vec{\nabla} \times \alpha \vec{B}_0) . \label{ii'}
\end{eqnarray}
Because the background fields and geometry are stationary and axisymmetric, 
the perturbed fields can have the form of $\exp(im \phi - i \sigma t)$, 
where $m$ is an arbitrary integer and $\sigma$ is an arbitrary complex 
number. The key point for analyzing the linear equations is to introduce 
the plasma displacement $\vec{\xi}$ defined by $\vec{v}_1 = - i \sigma 
\vec{\xi}$, for which we can give $\vec{\zeta} \equiv \vec{\xi} \times 
\vec{B}_{0}$. Then, from Eqs. (\ref{v'}) and (\ref{i'}), the perturbed 
fields $\vec{B}_1$ and $\vec{E}_1$ are explicitly solved as follows,
\begin{eqnarray}
\vec{B}_1 = \nu (\vec{\nabla} \times \alpha \vec{\zeta} \ ) + i 
\frac{(\vec{B}_{1}^{p} \cdot \vec{\nabla} \Omega_F)}{\sigma - m \Omega_F} 
\varpi\vec{e}_{T} ,
\label{B1} \\
\vec{E}_1 = \nu \frac{\varpi(\Omega_F - \omega)}{\alpha}\left[(\vec{\nabla} 
\times \alpha \vec{\zeta}) \times \vec{e}_{T} \right] + i \sigma \vec{\zeta} ,
\end{eqnarray}
where $\nu  \equiv \sigma/(\sigma - m \Omega_F)$.

For mathematical simplicity in later calculations, we assume that 
$\Omega_F$ is constant all over the field lines, i.e., the second term of 
the right-hand side of Eq. (\ref{B1}) vanishes. Then, by substituting 
$\vec{B}_1$ and $\vec{E}_1$ into Eq. (\ref{ii'}), our problem reduces to 
the eigenvalue problem for the eigenfrequency $\sigma$ and the eigenvector 
$\vec{\zeta}$ (or $\vec{\xi}$ \ ), which has the form
\begin{equation}
\sum_{n=0}^{2} \ \sigma^{n} (\hat{A}_{n}\vec{\xi} \ ) = 0 .
\end{equation}
Because we use the variational principle to consider the eigenvalue 
problem, here we express the differential operators $\hat{A}_{n}$ in terms 
of the integrals of arbitrary displacement vectors $\vec{\xi}$ and 
$\vec{\eta}$ as follows,
\begin{equation}
a_{n} \equiv \int d^{3}r \ \sqrt{h} \ \vec{\eta}^{\ *}\cdot
(\hat{A}_{n}\vec{\xi} \ ) \equiv \int d^{3}r \ \alpha\sqrt{h} \ b_{n} ,
\end{equation}
where stars and $h$ denote complex conjugates and the determinant of the 
metric $h_{ij}$, respectively. Some parts of the integrands
$\vec{\eta}^{\ *}\cdot(\hat{A}_{n}\vec{\xi} \ )$ reduce to the total 
divergence written by $\vec{\nabla}\cdot\vec{b}$, which is assumed to 
vanish owing to the condition such that no perturbations are generated at 
the boundaries. Then, through the lengthy calculations using $\vec{\zeta}$ 
and $\vec{\chi} \equiv \vec{\eta} \times \vec{B}_{0}$ instead of 
$\vec{\xi}$ and $\vec{\eta}$, we arrive at the results
\begin{eqnarray}
b_{2} &=& \vec{\chi}^{\ *}\cdot\vec{\zeta} , \\
b_{1} &=& -2m\omega\vec{\chi}^{\ *}\cdot\vec{\zeta}
+ i\frac{\alpha (\vec{\nabla} \cdot \vec{E}_{0})}{B_{0}^{2}}\vec{B}_{0}
\cdot(\vec{\chi}^{\ *} \times \vec{\zeta} \ ) \nonumber \\
&+& \frac{i}{\alpha}(\Omega_{F}-\omega) \left[ \vec{\chi}^{\ *} \cdot 
\vec{\nabla}(\varpi\alpha\zeta_{T})- \vec{\zeta} \cdot \vec{\nabla}
(\varpi\alpha\chi_{T}^{*}) \right] , \\
b_{0} &=& m^2\omega^{2}\vec{\chi}^{\ *}\cdot\vec{\zeta} + 
\frac{(\Omega_{F}-\omega)^{2}}{\alpha^{2}}\vec{\nabla}
(\varpi\alpha\chi_{T}^{*})\cdot\vec{\nabla}(\varpi\alpha\zeta_{T}) \nonumber \\
&-& i \frac{m\alpha\Omega_{F}(\vec{\nabla} \cdot \vec{E}_{0})}{B_{0}^{2}}
\vec{B}_{0}\cdot(\vec{\chi}^{\ *} \times \vec{\zeta} \ )
- \frac{im\omega}{\alpha}(\Omega_{F}-\omega) \left[ \vec{\chi}^{\ *} \cdot 
\vec{\nabla}(\varpi\alpha\zeta_{T}) - \vec{\zeta} \cdot \vec{\nabla}
(\varpi\alpha\chi_{T}^{*}) \right]  \nonumber \\
&-& (\vec{\nabla}\times\alpha\vec{\chi}^{\ *})\cdot
(\vec{\nabla}\times\alpha\vec{\zeta} \ )
- 2\pi \left( \frac{dI}{d\Psi} \right)
\left[ \vec{\chi}^{\ *}\cdot(\vec{\nabla} \times \alpha\vec{\zeta} \ )
+ \vec{\zeta}\cdot(\vec{\nabla} \times \alpha\vec{\chi}^{\ *}) \right] 
\end{eqnarray}
where $\zeta_{T}$ and $\chi_{T}$ are the toroidal components of 
$\vec{\zeta}$ and $\vec{\chi}$ measured in the orthonormal bases. These 
equations clearly show that the operators $\hat{A}_{n}$ are Hermitian.

If the vector $\vec{\chi}$ is chosen to be equal to $\vec{\zeta}$, the 
coefficients $a_{n}$ appearing in the dispersion relation
\begin{equation}
a_{2}\sigma^{2}+a_{1}\sigma+a_{0} = 0 \label{dis}
\end{equation}
become real numbers. Because $a_{2}$ is positive definite, it is clear that 
the sufficient and necessary condition for the screw instability of 
stationary configurations is
\begin{equation}
W \equiv W_{P}+W_{R} < 0,  \label{W}
\end{equation}
where $W_{P} \equiv -a_{0}$ is the so-called potential energy for the small 
displacement $\vec{\xi}$ \cite{kad66,kam67}. Even if $W_{P}$ is not 
positive semi-definite, stationary configurations in rotating 
magnetospheres can be stabilized by virtue of the existence of the 
rotational term $W_{R} \equiv a_{1}^{2}/4a_{2}$.

It should be remarked that $a_{n}$ are the functional of $\vec{\zeta}$ 
which is perpendicular to $\vec{B}_{0}$. This motivates us to introduce the 
poloidal orthonormal basis vectors defined by
\begin{equation}
\vec{e}_{\parallel} = \frac{\vec{B}_{0}^{p}}{|B_{0}^{p}|} ,  \ \
\vec{e}_{\bot} = \frac{\vec{\nabla} \Psi}{2 \pi \varpi |B_{0}^{p}|} ,
\end{equation}
in addition to the toroidal basis vector $\vec{e}_{T}$ with the condition 
$\vec{e}_{\parallel}=\vec{e}_{\bot}\times\vec{e}_{T}$. Then, decomposing 
the vector $\vec{\zeta}$ into
\begin{equation}
\vec{\zeta} = \zeta_{\parallel} \vec{e}_{\parallel} + \zeta_{\bot} 
\vec{e}_{\bot} + \zeta_{T}\vec{e}_{T} ,
\end{equation}
we obtain
\begin{equation}
\zeta_{T} = -\xi_{\bot}B_{0}^{p} , \ \ \zeta_{\parallel} = \xi_{\bot}B_{0}^{T}
\end{equation}
for $\vec{B}_{0} = B_{0}^{p}\vec{e}_{\parallel} + B_{0}^{T}\vec{e}_{T}$ and 
$\xi_{\bot}=\vec{e}_{\bot}\cdot\vec{\xi}$. Hence, we can treat the two 
functions $\xi_{\bot}$ and $\zeta_{\bot}$ as the trial functions to 
calculate the eigenvalue $\sigma$.

We have established the basic formulae to use the variational principle for 
the stability analysis in Kerr geometry. The next step will be to apply the 
method to Eq. (\ref{dis}) with the purpose of determining the trial 
functions and obtaining the instability criterion expressed by stationary 
background fields. Unfortunately, this becomes a very complicated task 
owing to the geometrical terms involved in $a_{n}$. Then, our interest in 
the next section is focused on a stabilizing effect of $\Omega_{F}$ in 
rotating magnetospheres, considering simple stationary fields and the 
perturbations given in flat spacetime.

\section{STABILIZING EFFECT IN ROTATING MAGNETOSPHERES}
\label{sec:level3}

As a mechanism of jet formation the existence of collimated field lines 
threading a black hole is astrophysically interesting. The jets would 
propagate along the poloidal magnetic field lines to large distances, 
keeping the collimated structure under an action of the toroidal magnetic 
field. Here, the stationary magnetic field $\vec{B}_{0}$ is assumed to be 
cylindrical as a typical configuration of collimated field lines. We 
discuss the instability of such a global structure due to the perturbations 
$\vec{\xi}$ with a wave length (in the vertical direction along the filed 
line) much longer than the horizon scale. (Long-wave modes are expected to 
become dominant in the screw instability \cite{kad66}.) Then, the integrals 
in the vertical direction appearing in $a_{n}$ should extend to a very 
large scale such that $L \gg M$, where gravity becomes very weak. This 
allows us to estimate approximately the integrals $a_{n}$ by giving fields 
in flat spacetime.

Because of the cylindrical configuration of the stationary magnetic field, 
it is convenient to use the cylindrical coordinates denoted by $\varpi$, 
$\phi$ and $z$, for which the Euclidean metric $h_{ij}$ is given by
\begin{equation}
h_{ij}dx^{i}dx^{j} = d\varpi^{2}+\varpi^{2}d\phi^{2}+dz^{2} ,
\end{equation}
and the orthonormal basis vectors previously introduced become
\begin{equation}
\vec{e}_{\parallel} = \vec{e}_{z}, \ \ \vec{e}_{\bot} = \vec{e}_{\varpi} .
\end{equation}
The magnetic components are assumed to be dependent on $\varpi$ only as 
follows,
\begin{equation}
\vec{B}_{0} = B(\varpi)\vec{e}_{z}+U(\varpi)\vec{e}_{T} ,    \label{b0}
\end{equation}
and for $\alpha=1$ and $\omega=0$ the Grad-Shafranov equation (\ref{GS}) 
reduces to
\begin{equation}
  (1-\varpi^2 \Omega_F^2) BB' + UU' = 2\varpi\Omega_F^2 B^2- 
\frac{1}{\varpi}U^2, \label{GS2}
\end{equation}
where primes denote the differentiation with respect to $\varpi$. To derive 
Eq. (\ref{GS2}), we have also used the relation
\begin{equation}
\frac{dI}{d\Psi} = -\frac{(\varpi U)'}{4\pi\varpi B}
\end{equation}
for the poloidal current $I$. If the toroidal component $U$ is given, we 
can determine the distribution of the poloidal component $B$. The simplest 
example giving a constant $B$ would be $U=\pm\varpi\Omega_{F}B$.

Now the perturbations with the eigenfrequency $\sigma$ can have the form
\begin{eqnarray}
\xi_{\bot} = \xi_{\varpi} = \xi(\varpi)\, \exp(i m \phi + i k z - i \sigma t) 
, \nonumber \\
\zeta_{\bot} = \zeta_{\varpi} = \zeta(\varpi)\, \exp(i m \phi + i k z - i 
\sigma t) ,
\end{eqnarray}
where $k$ is the wave number in the $z$-direction, and the minimal value of 
$k$ becomes approximately $L^{-1}$ in the magnetosphere with the scale $L$. 
To obtain a relation between the two trial functions $\xi$, $\zeta$, we 
consider the variation of Eq. (\ref{dis}). As will be seen later, the 
variational principle becomes consistent if $\vec{\eta}$ is chosen to be an 
eigenvector with the eigenvalue $\sigma^{*}$ (which may be complex), i.e.,
\begin{eqnarray}
\eta_{\bot} = \eta_{\varpi} = \eta(\varpi)\, \exp(i m \phi + i k z - i 
\sigma^{*} t) , \nonumber \\
\chi_{\bot} = \chi_{\varpi} = \chi(\varpi)\, \exp(i m \phi + i k z - i 
\sigma^{*} t) ,
\end{eqnarray}
where $\eta=\xi(\varpi,\sigma^{*})$ and $\chi=\zeta(\varpi,\sigma^{*})$.

Using these eigenvectors $\vec{\zeta}$ and $\vec{\chi}$, the coefficients 
$a_{n}$ in Eq. (\ref{dis}) reduce to
\begin{equation}
   a_{n} = 2\pi L \int \varpi d\varpi \ b_{n} .
\end{equation}
with $L$ as the length in the $z$-direction. It is easy to calculate the 
integrands $b_{2}$ and $b_{1}$ with the results
\begin{equation}
   b_{2} = \chi^{*}\zeta+(B^{2}+U^{2})\eta^{*}\xi  ,
\end{equation}
\begin{equation}
   b_{1} = i\Omega_{F} \left[ \zeta(\varpi B\eta'^{*}-B\eta^{*})-\chi^{*}
           (\varpi B\xi'-B\xi)+2i(mB-k\varpi U)B\eta^{*}\xi \right] .
\end{equation}
To calculate $b_{0}$, the partial integration of the terms containing 
$(\eta^{*}\xi)'$ becomes necessary. By omitting the contribution from the 
boundary, we have
\begin{equation}
b_{0} = 
   P\eta'^{*}\xi'+Q\eta^{*}\xi+\frac{m^{2}+k^{2}\varpi^{2}}{\varpi^{2}}
   \left( i\chi^{*}-\frac{S_{\eta}^{*}}{m^{2}+k^{2}\varpi^{2}} \right)
   \left( i\zeta+\frac{S_{\xi}}{m^{2}+k^{2}\varpi^{2}} \right) ,
\end{equation}
where
\begin{eqnarray}
P &=& -\frac{(mU+k\varpi B)^{2}}{m^{2}+k^{2}\varpi^{2}}
   +(\varpi\Omega_{F}B)^{2} , \\
Q &=& \frac{(mU+k\varpi B)^{2}}{\varpi^{2}}
           \left( \frac{1}{m^{2}+k^{2}\varpi^{2}}-1 \right)
     -\frac{2k^{2}(k^{2}\varpi^{2}B^{2}-m^{2}U^{2})}
           {(m^{2}+k^{2}\varpi^{2})^{2}} \nonumber \\
&-&\Omega_{F}^{2} \left[ (1-m^{2}-k^{2}\varpi^{2})B^{2} 
     -\frac{2k^{2}\varpi^{2}(\varpi BB'+2B^{2})}
           {m^{2}+k^{2}\varpi^{2}} \right] , \\
S_{\xi} &=& (mB-k\varpi U)\varpi\xi'+(mB+k\varpi U)\xi , \\
S_{\eta} &=& (mB-k\varpi U)\varpi\eta'+(mB+k\varpi U)\eta .
\end{eqnarray}
It should be noted that no derivatives of $\zeta$ and $\chi$ appears in the 
integrands. Then, taking the variation $\delta$ of Eq. (\ref{dis}) and 
putting $\delta\xi=\delta\eta^{*}=0$, we require that the coefficients of 
$\delta\zeta$ and $\delta\chi^{*}$ vanish, because they are algebraically 
independent and arbitrary. This procedure leads to the results
\begin{equation}
(m^{2}+k^{2}\varpi^{2}-\sigma^{2}\varpi^{2})\zeta = 
iS_{\xi}-i\Omega\sigma\varpi^{2}(\varpi B\xi'-B\xi) , \label{trial1}
\end{equation}
and
\begin{equation}
(m^{2}+k^{2}\varpi^{2}-\sigma^{2}\varpi^{2})\chi^{*} = 
  -iS_{\eta}^{*}+i\Omega\sigma\varpi^{2}(\varpi B\eta'^{*}-B\eta^{*}) ,
\end{equation}
while the latter is automatically satisfied by virtue of 
$\eta=\xi(\varpi,\sigma^{*})$ and $\chi=\zeta(\varpi,\sigma^{*})$.

Since $\zeta$ has been fixed by the variational method, the dispersion 
relation (\ref{dis}) is now given by
\begin{equation}
\int \varpi d\varpi ( F\eta'^{*}\xi'+G\eta^{*}\xi+H_{\sigma} ) = 0 , 
\label{dis2}
\end{equation}
where
\begin{equation}
F = -\frac{(mU+k\varpi B)^{2}}{m^{2}+k^{2}\varpi^{2}} ,
\end{equation}
\begin{equation}
G = \frac{(mU+k\varpi B)^{2}}
         {\varpi^{2}} \left( \frac{1}{m^{2}+k^{2}\varpi^{2}}-1 \right)
   -\frac{2k^{2}(k^{2}\varpi^{2}B^{2}-m^{2}U^{2})}
         {(m^{2}+k^{2}\varpi^{2})^{2}}-\frac{m^{2}\Omega_{F}^{2}
          (\varpi^{4}B^{2})'}{\varpi^{3}(m^{2}+k^{2}\varpi^{2})} ,
\end{equation}
and the term dependent on $\sigma$ is
\begin{eqnarray}
H_{\sigma} &=& 
  \frac{m^{2}+k^{2}\varpi^{2}}{m^{2}+k^{2}\varpi^{2}-\sigma^{2}\varpi^{2}} 
  \nonumber \\
&\times& \left[ \frac{\sigma S_{\eta}^{*}}{m^{2}+k^{2}\varpi^{2}}
  -\Omega_{F}B(\varpi\eta'^{*}-\eta^{*}) \right] 
   \left[ \frac{\sigma S_{\xi}}{m^{2}+k^{2}\varpi^{2}}
  -\Omega_{F}B(\varpi\xi'-\xi) \right] \nonumber \\
&+& \left[ (\sigma-m\Omega_{F})^{2}B^{2}+(\sigma U+k\varpi\Omega_{F}B)^{2} 
    \right] \eta^{*}\xi .
\end{eqnarray}
If $\sigma$ is real, the term $F\eta'^{*}\xi'$ is negative semi-definite. 
Then, if it becomes dominant in Eq. (\ref{dis2}), the third term 
$H_{\sigma}$ can be positive for giving consistently a real eigenvalue. 
This means that the most dangerous mode to make $\sigma$ complex is 
obtained by choosing $\xi$ in such a way that the integral involving the 
term $F\eta'^{*}\xi'$ vanishes. In fact, for $\Omega_{F}=0$ \cite{gru99}, 
the screw instability has been found to occur for the mode given by
\begin{equation}
\xi = \left\{
  \begin{array}{ll} 
  1 \ ,& \mbox{for} \ \ \varpi \leq \varpi_{0},                       \\
   \{\varpi_{0}(1+\mu)-\varpi\}/(\varpi_{0}\mu) \ ,
       & \mbox{for} \ \ \varpi_{0}\leq \varpi \leq \varpi_{0}(1+\mu), \\
  0 \ ,& \mbox{for} \ \ \varpi_{0}(1+\mu) \leq \varpi,
  \end{array}
  \right. \label{trial2}
\end{equation}
where $\varpi_{0}$ is the first zero of $mU+k\varpi B$ which is assumed to 
be present in the range $0<\varpi<\infty$. Taking the limit 
$\mu\rightarrow 0$, we can easily estimate the integral of 
$F\eta'^{*}\xi'$ to be of the order of $\mu$. (This trial function is 
defined without including $\sigma$, and so we have $\xi=\eta$.) We use the 
same trial function to analyze the dispersion relation (\ref{dis2}) 
extended to the case of $\Omega_{F}\neq 0$.

Let us assume that $B$ is everywhere positive. For regularity of the 
toroidal field on the symmetry axis, $U$ should approach zero at least in 
proportion to $\varpi$ in the limit $\varpi\rightarrow 0$. Further let the 
wave number $k$ be positive. Then, the sign of $m$ is chosen according to 
the requirement such that $mU<0$. By virtue of the steep gradient of $\xi$ 
at $\varpi=\varpi_{0}$, the integral of $H_{\sigma}$ containing the 
derivatives $\eta'^{*}\xi'$ becomes of the order of $1/\mu$ in 
contradiction to Eq. (\ref{dis2}), unless the eigenvalue $\sigma$ is 
suitably determined. It is easy to see that such a divergent term vanishes, 
if $\sigma$ is given by
\begin{equation}
\sigma = m\Omega_{F}+\sigma_{1} .
\end{equation}
The small correction $\sigma_{1}$ turns out to be of the order of 
$\mu^{1/2}$, because in the limit $\mu\rightarrow 0$ Eq. (\ref{dis2}) 
reduces to
\begin{equation}
\frac{m^{2}+k^{2}\varpi_{0}^{2}}{m^{2}
      (1-\Omega_{F}^{2}\varpi_{0}^{2})+k^{2} \varpi_{0}^{2}}
\frac{\varpi_{0}^{2}B_{0}^{2}}{m^{2}}
\frac{\sigma_{1}^{2}}{\mu} 
= \int_{0}^{\varpi_{0}} \frac{d\varpi}{\varpi} J , \label{dis3}
\end{equation}
where $B_{0} = B(\varpi_{0})$, and we have
\begin{eqnarray}
J &=& (mU+k\varpi B)^{2} \left[ 1-\Omega_{F}^{2}\varpi^{2}
     -\frac{1+\Omega_{F}^{2}\varpi^{2}}{m^{2}
(1 -\Omega_{F}^{2}\varpi^{2})+k^{2}\varpi^{2}} \right] \nonumber \\
&+& 
(m^{2}U^{2}-k^{2}\varpi^{2}B^{2})
\frac{2(m^{2}\Omega_{F}^{2}-k^{2})\varpi^{2}}
     { [ m^{2}(1-\Omega_{F}^{2}\varpi^{2})+k^{2}\varpi^{2} ]^{2} } .
\label{J1}
\end{eqnarray}
Though the eigenvalue of $\sigma_{1}$ is not determined unless the gradient 
$\mu^{-1}$ of $\xi$ at $\varpi=\varpi_{0}$ is precisely given, it is 
easy to see the sign of $\sigma_{1}^{2}$ from Eq. (\ref{dis3}). We note 
that the stability of stationary fields may drastically change if the light 
cylinder (i.e., $\Omega_{F}\varpi=1$) is present in the range 
$0<\varpi<\varpi_{0}$. However, our simple force-free model would not 
remain approximately valid in a region beyond the light cylinder where 
plasma inertia becomes very important. Hence, we would like to limit our 
stability analysis to the fields inside the light cylinder, i.e., 
$\varpi_{0}\Omega_{F}<1$, and we can conclude that the stationary fields 
given by $B$ and $U$ become unstable if the condition
\begin{equation}
\int_{0}^{\varpi_{0}} \frac{d\varpi}{\varpi} J < 0 \label{criterion}
\end{equation}
is satisfied. This is equivalent to Eq. (\ref{W}), if $W$ is estimated by 
using the eigenfrequency $\sigma=\Omega_{F}$ and the trial functions given 
by Eqs. (\ref{trial1}) and (\ref{trial2}).

 From the first term proportional to $(mU+k\varpi B)^{2}$ in $J$ the screw 
mode with $|m|=1$ (and $mU=-|U|$) turns out to be most important for the 
instability. To compare clearly the instability condition with the usual 
Kruskal-Shafranov criterion, let both $k$ and $\Omega_{F}$ be positive and 
small, i.e., $k\varpi_{0}\ll 1$ and $\Omega_{F}\varpi_{0}\ll 1$. For the 
long-wave screw mode giving $mU=-|U|$ in slowly rotating magnetospheres we 
have
\begin{equation}
J = -(|U|-k\varpi B)(k^{2}+\Omega_{F}^{2})\varpi^{2} 
    \left[ |U|-k\varpi B 
      +\frac{4(k^{2}-\Omega_{F}^{2})}{k^{2}+\Omega_{F}^{2}}k\varpi B 
    \right]                                                     \label{J2}.
\end{equation}
Then, if $\Omega_{F}<k$, it is clear that the requirement $|U|>k\varpi B$ 
in the range $0<\varpi<\varpi_{0}$ makes $J$ negative. Because the minimal 
value of the wave number $k$ is of the order of $L^{-1}$, we arrive at the 
Kruskal-Shafranov criterion
\begin{equation}
\frac{|U|}{\varpi} > \frac{B}{L}  \label{KS}
\end{equation}
as the instability condition for a twisted magnetic tube with the width 
$\varpi$. The important point obtained here is that the minimal value of 
$k$ is also restricted by $\Omega_{F}$, and we can only claim that the 
screw instability surely occurs if the condition
\begin{equation}
\frac{|U|}{\varpi} > \Omega_{F}B  \label{inst}
\end{equation}
is also satisfied.

If the condition (\ref{inst}) breaks down, the stability becomes a more 
subtle problem depending on the distribution of $B$ and $U$ as functions of 
$\varpi$. For example, let us consider a stationary model such that the 
poloidal field is nearly uniform ($B\simeq B_{0}$), while the weak toroidal 
field (i.e., $|U| \ll B$) is given by $|U|=\Omega_{0}\varpi B_{0}$ in the 
range $0<\varpi<\varpi_{1}$ and $|U|=\Omega_{0}\varpi_{1} B_{0}$ in the 
range $\varpi_{1}<\varpi$. Though the weak dependence of $B$ on $\varpi$ 
may be calculated from Eq. (\ref{GS2}), it is not important when we 
estimate the integral of $J$. The first zero of $|U|-k\varpi B$ appears at 
$\varpi=\varpi_{0}=\Omega_{0}\varpi_{1}/k$, and the condition $|U|>k\varpi 
B$ in the range $0<\varpi<\varpi_{0}$ leads to $k<|\Omega_{0}|$. Then, 
through the calculation of the integral of $J$, we can easily show that the 
stationary model with $|\Omega_{0}|\leq\Omega_{F}$ becomes stable even for 
the screw mode satisfying the Kruskal-Shafranov criterion.

However, one may consider a different model with the nearly uniform 
poloidal field and the weak toroidal field given by $|U|=\Omega_{0}\varpi 
B_{0}\left[ 1-(\varpi/\varpi_{2})^{2} \right]^{1/2}$ in the range 
$0<\varpi<\varpi_{2}$ and $|U|=0$ in the range $\varpi_{2}<\varpi$. Then, 
if the first zero $\varpi_{0}$ is chosen to be very close to $\varpi_{2}$, 
we have $|U(\varpi)| \gg |U(\varpi_{0})| = k\varpi_{0}B > k\varpi B$ (i.e., 
$J \simeq -U^{2}\Omega_{F}^{2}\varpi^{2} < 0$) in the range 
$0<\varpi<\varpi_{0}$, except near $\varpi=\varpi_{0}$. The stabilizing 
effect of $\Omega_{F}$ becomes very weak for the stationary model with 
$|U|$ which rapidly decreases as $\varpi$ increases.

The conclusion that the latter model with a decreasing $|U|$ is 
screw-unstable will be premature when the toroidal field vanishes at 
$\varpi=\varpi_{2}$ very close to the light cylinder (i.e., 
$\Omega_{F}\varpi_{2}\simeq 1$). This is because we must use Eq. (\ref{J1}) 
for $J$, when $k$ is chosen to be very small according to the requirement 
$\varpi_{0}\simeq\varpi_{2}$. Now we obtain the approximate form
\begin{equation}
J \simeq -U^{2}\Omega_{F}^{2}\varpi^{2}
        \frac{1-4\Omega_{F}^{2}\varpi^{2}+\Omega_{F}^{4}\varpi^{4}}
             {(1-\Omega_{F}^{2}\varpi^{2})^{2}} ,
\end{equation}
except near $\varpi=\varpi_{0}$, and it is easy to see that the integral of 
$J$ becomes positive by virtue of $\Omega_{F}\varpi_{0}\simeq 1$. Thus, we 
can expect that the rotating field is well stabilized against the screw 
mode with $k$ smaller than $\Omega_{F}$, unless a significant decrease of 
the toroidal field occurs in an inner region distant from the light 
cylinder.

In summary, we have presented the new criterion (\ref{criterion}), in which 
$J$ is given by Eq.(\ref{J2}) for long-wave screw modes in slowly rotating 
magnetospheres. If the more precise form (\ref{J1}) of $J$ is used, we can 
show a strong stabilizing effect due to the existence of the light 
cylinder. It is sure that the Kruskal-Shafranov criterion (\ref{KS}) is not 
a sufficient condition for the screw instability if the twisted magnetic 
tube is rotating. Importantly, in rotating magnetospheres with 
$\Omega_{F}>L^{-1}$, the toroidal magnetic field may be stably amplified 
to the maximum value given by $|U|=\Omega_{F}\varpi B$, instead of 
$|U|=\varpi B/L$. 
So, we can say that the stationary magnetic field  (\ref{b0}) is 
quite a reasonable assumption as a global field structure for the jets 
except local structures as a knot. 
If the deviation from the cylindrical field is serious, the criterion 
based on the integral (\ref{criterion}) would not be simply applicable. 
However, by rough estimations, we can expect that the stability effect 
by $\Omega_F$ remains valid in modified criterion.  
This will be crucial for discussing highly energetic phenomena in 
rotating magnetospheres.

\section{DISCUSSION}
\label{sec:level4}

Now we would like to give a brief comment on the application of the results 
obtained in the previous section to black hole magnetospheres. Let us 
assume that radial magnetic field lines thread a Kerr black hole which is 
slowly rotating with the angular velocity $\omega_{H}$. The split monopole 
magnetic field \cite{BZ77} is given by
\begin{equation}
\Psi = -2\pi c_{0}\; \cos\theta, \ \ B_{0}^{T} = 
  c_{0}(\Omega_{F}-\omega_{H})\; \sin^{2}\theta \label{radial}
\end{equation}
with a constant $c_{0}$, from which we have
\begin{equation}
B_{0}^{T} \simeq (\Omega_{F}-\omega_{H})\varpi B_{0}^{p}
\end{equation}
in the region distant from the event horizon. As $r$ increases, the radial 
magnetic field lines would tend to be collimated toward a cylindrical 
configuration, conserving the magnetic flux $\varpi^{2}B_{0}^{p}$ and the 
electric current $\varpi B_{0}^{T}$. Then, the stationary cylindrical 
field, for which the instability condition was discussed, would satisfy the 
relation
\begin{equation}
U \simeq (\Omega_{F}-\omega_{H})\varpi B , \label{UB}
\end{equation}
for the field lines threading the central black hole, i.e., in a limited 
range such that $0<\varpi<\varpi_{1}$.

The constraint on the angular velocity $\Omega_{F}$ of the magnetic field 
lines has been studied in \cite{li00} using the Kruskal-Shafranov 
criterion. However, if the length $L$ of the magnetic tube is larger than 
$\Omega_{F}^{-1}$, the sufficient condition of the screw instability should 
be replaced by Eq. (\ref{inst}), which leads to
\begin{equation}
|\Omega_{F}-\omega_{H}| > \Omega_{F} ,
\end{equation}
and we obtain
\begin{equation}
\Omega_{F} < \omega_{H}/2 .
\end{equation}
If the magnetic field rotates too slowly (see, e.g., \cite{pun98} for such 
a black-hole driven wind model), the toroidal field generated by the black 
hole becomes too strong to be stably sustained against the screw modes. 
This may yield a flaring event in plasma jets, because the stored magnetic 
energy is released to the kinetic energy of surrounding plasma \cite{gru99}.

If the magnetic field lines threading the black hole satisfy the condition
\begin{equation}
\Omega_{F} \geq \omega_{H}/2 ,
\end{equation}
a stable configuration becomes possible. In particular, the 
Blandford-Znajek mechanism \cite{BZ77} can work as a steady process of 
energy extraction from the rotating black hole, if $\Omega_{F}$ is in the 
range $\omega_{H}/2 \leq \Omega_{F} < \omega_{H}$. Interestingly, the 
magnetic field can be marginally stable, when the power becomes maximum, 
i.e., $\Omega_{F} = \omega_{H}/2$. Thus, the jets can be significantly 
powered by the Blandford-Znajek mechanism at large distances $L$ far from 
the central black hole, contrary to the claim in \cite{li00}.

The force-free model (\ref{radial}) is based on the boundary condition at 
the event horizon, which may not be justified if the effect of plasma 
accretion is taken into account. If we relax the constraint at the event 
horizon, the relation (\ref{UB}) would not always hold. Though the 
condition $\Omega_{F} = \omega_{H}/2$ of the marginal stability could be 
modified, it is plausible that the ratio $\omega_{H}/\Omega_{F}$ is a 
crucial parameter for determining a dominant energy source which powers 
jets produced in black hole magnetospheres: The screw instability will 
yield a helical structure of plasma jets and flare-like emission through 
the processes such as shock waves, while the Blandford-Znajek mechanism 
will provide steady Poynting flux for plasma acceleration. The physical 
mechanism to determine uniquely the parameter $\omega_{H}/\Omega_{F}$ is 
also controversial \cite{PC90}. If various values of the ratio 
$\omega_{H}/\Omega_{F}$ are permissible for rotating black hole 
magnetospheres in AGNs, such two different types of jets may be observable.

There exists also the Newtonian solution of Eq. (\ref{GS}) for a rotating, 
paraboloidal and disk-connected field \cite{Bla76}, satisfying the 
condition $B_{0}^{T} = - \Omega_{F}\varpi B_{0}^{p}$ at large distances. If 
the same instability condition holds even for the paraboloidal field, such 
a stationary model would be marginally stable. In this paper, we have 
considered magnetic field lines connecting a central black hole with remote 
astrophysical loads, in relation to jet formation. As was suggested in 
\cite{gru99}, closed field lines connecting the black hole 
with a surrounding disk may also become screw-unstable (see \cite{tom01} 
for a non-rotating model). Thus, the  
analysis of the screw instability of various stationary fields including 
the effect of strong gravity is a necessary task for understanding its 
astrophysical significance in rotating black hole magnetospheres,  
and the basic formulae discussed in Sec.II for the stability 
analysis in Kerr geometry should be fully applied. The results of 
this paper would be very useful as the foundation of the future problems. 

\end{document}